\documentclass[12pt]{iopart}
\begin{document}

\title[Reductions of integrable lattices]{Reductions of integrable lattices}

\author{A K Svinin}

\address{Institute for System
Dynamics and Control Theory, Siberian Branch of
Russian Academy of Sciences}
\ead{svinin@icc.ru}
\begin{abstract}
Based on the notion of Darboux-KP chain hierarchy and its invariant submanifolds we  construct some class of constraints compatible with integrable lattices. Some simple examples are given.
\end{abstract}

\maketitle

\section{Introduction}

Our concern in the paper is with differential-difference systems (lattices) over finite number of fields (unknown functions of discrete variable $i$) which share the property of having infinite number of conservation laws. In a sense one can interpret these conservation laws as analog of first integrals for finite-dimensional dynamical systems. It is supposed that to each conservation law corresponds the flow on suitable infinite-dimensional phase-space and these flows are pair-wise commuting. Differential-difference systems of this type are reffered to as `integrable'. For integrable lattice under consideration corresponding flows governed by some evolutionary equations can be interpreted as generalized symmetries. One says that given system with these properties admits integrable hierarchy (see, for example \cite{oevel}).

It is known from the literature a number of differential-difference systems with above-mentioned properties which have applications in different areas of natural sciences. Perhaps simplest and from the other hand interesting example of integrable lattice is well-known Volterra \cite{volterra} (or Kac-van Moerbeke) equation \cite{kac}.
\begin{equation}
a^{\prime}_i=a_i\left(a_{i-1}-a_{i+1}\right).
\label{Vl}
\end{equation}
It can be considered as single equation on function $a=a(i,x)=a_i$ of discrete variable $i\in{\bf Z}$ and continuous variable $x\equiv t_1\in{\bf R}$ and is known to be integrable discretization of Korteweg-de Vries equation \cite{kac}. Due to numerous applications of Volterra lattice (see, for example, \cite{fokas}, \cite{manakov}, \cite{volkov}), it can be accepted nowadays as classical equation of mathematical physics. This equation is known to be integrable by inverse scattering transform method due to \cite{manakov}. 

Volterra lattice hierarchy can be written as infinite number of evolutionary equations of the form
\begin{equation}
\partial_sa_i=a_i\left(\zeta_s(i+1)-\zeta_s(i-1)\right),\;\;\mbox{where}\;\;\partial_s\equiv \frac{\partial}{\partial t_s}.
\label{Vh}
\end{equation}
It turns out that a polynomial discrete functions $\zeta_k=\zeta_k[a]$ here are conservation densities for all the flows. One can write differential-difference conservation laws for (\ref{Vh}) as
\[
D_{t_s}\zeta_k(i)=J_{s,k}(i+1)-J_{s,k}(i).
\]
The discrete functions $\zeta_k$ can be calculated making of use recursion relation
\[
\Delta\zeta_{k+1}=L\zeta_k,\;\;\mbox{with}\;\;\zeta_1=-a
\]
which can be derived from Lax representation of Volterra lattice hierarchy (see, for example, \cite{vekslerchik}). This relation is defined by a pair of discrete operators: forward difference operator $\Delta\equiv\Lambda-1$ and $L\equiv (a\Lambda+a)\circ\Lambda^{-1}-\Lambda\circ(a\Lambda+a)$ whith $\Lambda$ being a shift operator acting on arbitrary function of discrete variable as $(\Lambda f)(i)=f(i+1)$.

Formally, one can construct first integrals of (\ref{Vh}) as $\zeta_k=\sum_{i\in{\bf Z}}\zeta_k(i)$ but, generally speaking, it makes no sense because of convergence problem. Provided periodicity condition $a_{i+N}=a_i$ which are evidently compatible (since right-hand side of (\ref{Vh}) does not depend explicitly on $i$) with all flows (\ref{Vh}) one is forced to consider finite-dimensional dynamical system 
\[
u_k^{\prime}=u_k(u_{k-1}-u_{k+1})\;\;\mbox{where}\;\; k\in{\bf Z}/N{\bf Z}.
\]  
It is known to be Liouville-integrable Hamiltonian system. 

One knows a class of constraints compatible with Volterra lattice which characterize similarity solutions of the latter and its hierarchy. Simplest example is given by the first discrete Painlev\'e equation \cite{fokas}
\begin{equation}
a_i\left(a_{i-1}+a_i+a_{i+1}-x\right)+\alpha_i=0.
\label{dPI}
\end{equation}
Here it is supposed that $\alpha_i$'s do not depend on $x$ and satisfy the relation $\alpha_{i+2}=\alpha_i+1$. The restriction of Volterra lattice (\ref{Vl}) with the help of (\ref{dPI}) leads to the system of ordinary differential equations
\begin{equation}
\begin{array}{l}
u_1^{\prime}=-u_1\left(u_1+2u_2-x\right)-a,
\\[0.2cm]
u_2^{\prime}=u_2\left(2u_1+u_2-x\right)+b,
\end{array}
\label{de}
\end{equation}
where $u_1=a_i, u_2=a_{i+1}$ and $a=\alpha_i, b=\alpha_{i+1}$ with some $i=i_0$. This system, as is known, is Hamiltonian and equivalent to fourth Painlev\'e equation integrable via isomonodromy method \cite{milne}. Compatibility of (\ref{dPI}) with Volterra lattice means that shifting $i\rightarrow i+1$ yields discrete symmetry transformation
\[
\bar{u}_1=u_2,\;\;\;
\bar{u}_2=-u_1-u_2-\frac{b}{u_2}+x,\;\;\;
\bar{a}=b,\;\;\;
\bar{b}=a+1
\]
for (\ref{de}).

It turns out that there exists an infinite number of conditions compatible with Volterra lattice hierarchy which can be written as ordinary autonomous $N$th-order difference equation
\begin{equation}
a_{i+N}=R(a_i,..., a_{i+N-1})
\label{aiN}
\end{equation}
with right-hand side $R$ being some rational function of its arguments. As a result we are led to the system of ordinary differential equations
\[
u_1^{\prime}=u_1(S(u_1,..., u_N)-u_2),
\] 
\begin{equation}
u_k^{\prime}=u_k(u_{k-1}-u_{k+1}),\;\;\;k=2,..., N-1
\label{ode}
\end{equation}
\[
u_N^{\prime}=u_N(u_{N-1}-R(u_1,..., u_N))
\]
on $u_1=a_i,..., u_N=a_{i+N-1}$ with some initial value $i=i_0$. The function $S$ here is defined through inversion formula $a_{i-1}=S(a_i,..., a_{i+N-1})$. Shifting $i\rightarrow i+1$ yields discrete symmetry transfomation 
\[
\bar{u}_k=u_{k+1},\;\; k=1,..., N-1, 
\]
\[
\bar{u}_N=R(u_1,..., u_N)
\]
for (\ref{ode}). Some examples of reductions generated by a kind of relations (\ref{aiN}) were presented in \cite{svinin3}. 

It is our main goal in this paper is to present unified geometric approach for constructing some class of restrictions compatible with integrable lattices and its hierarchies. We consider a community of integrable lattices which are supposed to be derived as a result of suitable reduction of the so-called Darboux-KP (DKP) chain hierarchy which, in fact, is a chain of KP hierarchy solutions related with each other by Darboux map \cite{mpz}. From geometric point of view it is convenient to present DKP chain in the form of differential and differential-difference conservation laws. One writes its equations in terms of infinite collections  $\{h_k, a_k\}$ of functions of evolution parameters $t_s$ and discrete variable $i$. It is relevant in this approach to consider these unknown functions as coefficients of some formal Laurent series.  

DKP chain hierarchy admits an infinite number of invariant submanifolds ${\cal S}_{l-1}^n$ and Volterra equation and many other integrable lattices arise as a result of restriction of DKP chain on submanifold constructed as an intersection ${\cal S}^n_0\cap{\cal S}_{l-1}^p$ with some $n ,p$ and $l$. In particular ${\cal S}_0^1\cap{\cal S}_0^2$ corresponds to Volterra lattice. It is important that all integrable lattices under consideration in this paper may be written in terms of universal coordinates $\{a_k=a_k(i) : k\geq 1\}$ parametrizing the points of some infinite-dimensional phase-space ${\cal M}$. Restricting DKP chain only on ${\cal S}_0^n$ yields a hierarchy which is more convenient to write in terms of infinite number of functions $\{a_k\}$. This hierarchy we call $n$th discrete KP, since in the case $n=1$ one has equations of ordinary discrete KP hierarchy. Integrable lattices over finite number of fields can be considered as a result of restriction of ${\cal M}$ to some submanifold ${\cal M}_{n,p,l}$ with the help of algebraic equations 
\[
I_k[a_1, a_2,...]=0,\;\;\mbox{with}\;\; k\geq 1 
\]      
where $I_k$' are some suitable polynomial discrete functions on ${\cal M}$. 

To make the matter more clear, let us illustraite our general result stated below in Theorem 3 by discussing simple example. Consider one-field lattice
\[
a_i^{\prime}+a_{i+1}^{\prime}=\left(a_i+a_{i+1}\right)\left(a_{i-1}-a_{i+2}\right)
\]  
which corresponds to ${\cal S}_0^1\cap{\cal S}_0^3$ and require that $a$ is also solution to Volterra equation. It is equivalent to condition
\begin{equation}
a_ia_{i+2}=a_{i+1}a_{i+3}
\label{condition}
\end{equation}
which, as can be checked by direct calculations, is compatible with Voltera lattice. At the first sight it seems that Volterra equation supplemented by constraint (\ref{condition}) corresponds to triple intersection ${\cal S}_0^1\cap{\cal S}_0^2\cap{\cal S}_0^3$, but one can check that it gives only trivial solution $a=0$ of it. The point is that there exist invariant conditions weaker than those defining ${\cal M}_{1,3,1}$. It turns out that this conditions are written in the form of periodicity relations on some discrete polynomial functions. The submanifold defined by this `weak' conditions we denote as ${\cal N}_{1,3,1}$ and Volterra equation supplemented by (\ref{condition}) appear as a result of restriction of discrete KP hierarchy on ${\cal M}_{1,2,1}\cap{\cal N}_{1,3,1}$    

This paper is organized as follows: in Section \ref{sec:DKP}, we provide the reader by some basic facts about DKP chain hierarchy following along the lines suggested in \cite{mpz}. In Section \ref{sec:lat}, we formulate theorem which provide us by an infinite number of invariant submanifolds for DKP chain hierarchy. This result is the basis to establish relationship of many integrable lattices with the KP hierarchy. We provide the reader by some examples of integrable differential-difference systems over finite number of fields. In Section \ref{sec:main} we formulate our main result which allows to construct a broad class of constraints compatible with integrable lattices and show, in Section \ref{sec:Bl}, how this result can be applied  on example of extended Volterra equation \cite{narita} 
\begin{equation}
a_i^{\prime}=a_i\left(\sum_{j=1}^na_{i-j}-\sum_{s=1}^na_{i+j}\right)
\label{Bl}
\end{equation}
known also as Bogoyavlenskii lattices \cite{bogoyavlenskii}. 

\section{DKP chain hierarchy}
\label{sec:DKP}

Let us first to give some basics on DKP chain \cite{mpz}. This can be defined through two relations
\begin{eqnarray}
\partial_sh(i)&=&\partial H^{(s)}(i),
\label{1}
\\
\partial_sa(i)&=&a(i)\left(H^{(s)}(i+1)-H^{(s)}(i)\right)
\label{2}
\end{eqnarray}
first of which yields evolution equations of KP hierarchy in the form of local conservation laws with $h=z+\sum_{k\geq 2}h_kz^{-k+1}$ being a generating function for conserved densities of KP hierarchy. The formal Laurent series
\[
H^{(s)}=z^s+\sum_{k\geq 1}H_k^sz^{-k},
\]
attached to any integer $s\geq 2$, is the generating function for suitable fluxes and can be uniquely defined as projection of $z^s$ on the space
\[
{\cal H}_{+}=<1, h, h^{(2)},...>
\]
spanned by Fa\`a di Bruno iterates $h^{(k)}\equiv (\partial
+h)^{k}(1)$. For instance, one has
\begin{equation}
H^{(1)}=h,\;\;\;
H^{(2)}=h^{(2)}-2h_2,\;\;\;H^{(3)}=h^{(3)}-3h_2h-3h_3-3h_2^{\prime},...
\label{forin}
\end{equation}
Thus the coefficients $H_k^s$ are defined as differential polynomials of $h_2, h_3,...$ Linear relations (\ref{forin}) are invertible and this means that any element of ${\cal H}_{+}$ can be written as suitable linear combination over $H^{(k)}$.

Representation of the KP hierarchy in the form (\ref{1}) is equivalent to Sato's formulation of that on the level of Lax equation
\[
\partial_s{\cal Q}=[\left({\cal Q}^s\right)_{+}, {\cal Q}],
\]
on formal pseudodifferential operator ${\cal Q}=\partial+\sum_{k\geq 2}u_k\partial^{-k+1}$. One has the following:
\[
\partial={\cal Q}+\sum_{k\geq 2}h_k{\cal Q}^{-k+1}
\]
and
\[
\left({\cal Q}^s\right)_{+}={\cal Q}^s+\sum_{k\geq 2}H_k^s{\cal
Q}^{-k+1}.
\]
One can write, for example, the following:
\[
h_2=-u_2,\;\; h_3=-u_3,\;\;\; h_4=-u_4-u_2^2,...
\]
and
\[
H_1^2=2h_3+h_2^{\prime}=-2u_3-u_2^{\prime},\;\;\; H_2^2=2h_4+h_3^{\prime}+h_2^2=-2u_4-u_3^{\prime}-u_2^2,...
\]

The relation (\ref{2}) on formal Laurent series $a(i)=z+\sum_{k\geq 1}a_k(i)z^{-k+1}$ relate two neighbours $h(i)$ and $h(i+1)$ by Darboux map $h(i)\rightarrow h(i+1)=h(i)+a_x(i)/a(i)$ and garantee compatibility of the latter with KP flows. It is obvious that this equation can be rewritten
in the form of differential-difference conservation laws
\[
\partial_s\xi(i)=H^{(s)}(i+1)-H^{(s)}(i)
\]
with
\begin{eqnarray}
\displaystyle
\fl
\xi&=&\ln a=\ln z +
\sum_{k\geq 1}a_kz^{-k}-
\frac{1}{2}\left(\sum_{k\geq 1}a_kz^{-k}\right)^2
+\frac{1}{3}\left(\sum_{k\geq 1}a_kz^{-k}\right)^3-\cdot\cdot\cdot
\nonumber \\
\fl
&\equiv&\ln z+\sum_{k\geq 1}\xi_kz^{-k} \nonumber
\end{eqnarray}
Thus, we consider following equations of DKP chain hierarchy
\begin{equation}
\partial_sh_k(i)=\partial H^s_{k-1}(i),\;\;\;
\partial_s\xi_k(i)=H^s_k(i+1)-H^s_k(i).
\label{DKP}
\end{equation}

\section{Integrable lattices}
\label{sec:lat}

\subsection{Invariant submanifolds of DKP chain}

As was mentioned in Introduction one knows from the literature many examples of differential-difference systems over finite number of fields which share the property of having infinite number of conservation laws and corresponding one-parametric generalized symmetry groups defined by respective evolutionary equations. There are different methods for constructing integrable lattices and its explicit solutions, such that Lax pairs, recursion operators etc. For consulting, see, for example, \cite{belov}, \cite{blaszak}, \cite{bogoyavlenskii}, \cite{kac}, \cite{kupershmidt}, \cite{narita}, \cite{oevel}, \cite{shabat}, \cite{toda}. 

In Refs. \cite{svinin1} and \cite{svinin2} we have proved  that DKP chain is a convenient and simple notion to show relationship of integrable lattices with KP hierarchy. This relationship is very useful due to remarkable Sato theory which gives description of analitic solutions of KP hierarchy in terms of infinite Grassmanian manifold and $\tau$-function \cite{ohta}. Following two theorems give a framework for constructing integrable lattices whose hierarchies directly related with KP hierarchy.

{\bf Theorem 1.}
\cite{svinin1} The submanifold ${\cal S}_{l-1}^n$ defined by condition
\begin{equation}
z^{l-n}a^{[n]}(i)\in{\cal H}_{+}(i),\;\; \forall
i\in{\bf Z} \label{sln}
\end{equation}
is tangent with respect to DKP chain flows defined by (\ref{DKP}).

{\bf Theorem 2.} \cite{svinin2} The chain of inclusions
of invariant submanifolds
\[
{\cal S}_{l-1}^n\subset {\cal S}_{2l-1}^{2n}\subset{\cal
S}_{3l-1}^{3n}\subset\cdot\cdot\cdot\subset {\cal
S}_{kl-1}^{kn}\subset\cdot\cdot\cdot
\]
is valid.

Here, by definition
\[
a^{[s]}(i)=\left\{
\begin{array}{l}
a(i)\times\cdot\cdot\cdot\times a(i+s-1),\;\;\; s\geq 1 \\
1,\;\;\; s=0  \\
a^{-1}(i-1)\times\cdot\cdot\cdot\times a^{-1}(i-|s|),\;\;\; s\leq
-1
\end{array}
\right.
\]
are discrete Fa\`a di Bruno iterates of Laurent series $a(i)$. In what follows, we use simple obvious identity
\begin{equation}
a^{[r_1+r_2]}(i)=a^{[r_1]}(i)a^{[r_2]}(i+r_1)=a^{[r_2]}(i)a^{[r_1]}(i+r_2),
\;\;\; \forall r_1, r_2\in{\bf Z}.
\label{s1s2}
\end{equation}
Using the coefficients $a_k^{[r]}$ defined through the relation
$
a^{[k]}=z^k+\sum_{s\geq 1}a_s^{[k]}z^{k-s}
$
one recovers from (\ref{s1s2}) 
\begin{eqnarray}
\displaystyle
\fl
a_k^{[r_1+r_2]}(i)&=&
a_k^{[r_1]}(i)+\sum_{j=1}^{k-1}a_j^{[r_1]}(i)a_{k-j}^{[r_2]}(i+r_1)+a_k^{[r_2]}(i+r_1)
\nonumber \\
\fl
&=&a_k^{[r_2]}(i)+\sum_{j=1}^{k-1}a_j^{[r_2]}(i)a_{k-j}^{[r_1]}(i+r_2)+a_k^{[r_1]}(i+r_2)
\;\;\; \forall r_1, r_2\in{\bf Z}.
\label{s1s21}
\end{eqnarray}

We observe that the condition (\ref{sln}) can be written in the form of the following generating relation:
\[
z^{l-n}a^{[n]}=H^{(l)}+\sum_{k=1}^la_k^{[n]}H^{(l-k)}.
\]
After a look at negative powers of $z$ in the latter formula, one gets explicit form of (\ref{sln}) as follows:
\[
G_k^{(n,l)}\equiv a^{[n]}_{k+l}-H_k^l-\sum_{j=1}^{l-1}a_j^{[n]}H_k^{l-j}=0,
\;\; \forall k\geq 1.
\]

\subsection{Restriction of DKP chain on ${\cal S}_0^n$}

Let us consider the case $l=1$ corresponding to the invariant submanifold ${\cal S}_0^n$. It is defined by conditions
\[
G_k^{(n,1)}\equiv a^{[n]}_{k+1}-H_k^1=0,
\;\; \forall k\geq 1.
\]
So, on ${\cal S}_0^n$ one has $H_k^1=h_{k+1}=a^{[n]}_{k+1}$. Theorem 2 says that
\[
{\cal S}_0^n\subset {\cal S}_1^{2n}\subset{\cal S}_2^{3n}\subset\cdot\cdot\cdot\subset {\cal S}_{k-1}^{kn}\subset\cdot\cdot\cdot
\]
and, hence,
\[
G_k^{(n,1)}=0\;\; \Rightarrow\;\; G_k^{(2n,2)}=0\;\; \Rightarrow\;\; G_k^{(3n,3)}=0\;\; \Rightarrow\;\; \cdot\cdot\cdot
\]
Solving successively these relations in favor of $H_k^s$ gives
\begin{equation}
H^s_k=F_k^{(n,s)}[a_1, a_2,...]\equiv a_{k+s}^{[sn]}+\sum_{j=1}^{s-1}q_j^{(n, sn)}a_{k+s-j}^{[(s-j)n]},
\label{Hps}
\end{equation}
where $q_k^{(n,r)}=q_k^{(n,r)}[a_1, a_2,...]$ are polynomial discrete functions defined through the relation
\begin{equation}
z^r=a^{[r]}+\sum_{k\geq 1}q_k^{(n,r)}z^{k(n-1)}a^{[r-kn]}
\label{zr}
\end{equation}
or, more exactly,
\begin{equation}
a_k^{[r]}+\sum_{j=1}^{k-1}a^{[r-jn]}q_j^{(n,r)}+q_k^{(n,r)}=0,\;\;\;
\forall k\geq 1.
\label{exact}
\end{equation}
Below the first few $q_k^{(n,r)}$ are written 
\[
q_1^{(n, r)}=-a_1^{[r]},\;\;\; q_2^{(n, r)}=-a_2^{[r]}+a_1^{[r]}a_1^{[r-n]},\;\;\;
\]
\[
q_3^{(n, r)}=-a_3^{[r]}+a_1^{[r]}a_2^{[r-n]}+a_1^{[r-2n]}a_2^{[r]}-a_1^{[r]}a_1^{[r-n]}a_1^{[r-2n]}.
\]
It can be shown that the functions $q_k^{(n, r)}$ are related with each other by the relation
\begin{eqnarray}
\displaystyle
\fl
q_1^{(n, r_1+r_2)}(i)&=&
q_k^{(n, r_1)}(i)+\sum_{j=1}^{k-1}q_j^{(n, r_1)}(i)q_{k-j}^{(n, r_2)}(i+r_1-jn)+q_k^{(n, r_2)}(i+r_1)
\nonumber \\
\fl
&=&q_k^{(n, r_2)}(i)+\sum_{j=1}^{k-1}q_j^{(n, r_2)}(i)q_{k-j}^{(n, r_1)}(i+r_2-jn)+q_k^{(n, r_1)}(i+r_2).
\label{r1r2}
\end{eqnarray}
Further, there is a need in more general than (\ref{exact}) relation
\begin{equation}
a_k^{[r]}(i)+\sum_{j=1}^{k-1}a_{k-j}^{[r-jn]}(i)q_j^{(n,r-p)}(i+p)+q_k^{(n,r-p)}(i+p)=a_k^{[p]}(i)
\label{integers}
\end{equation}
with any integers $r$ and $p$. The latter can be easily obtained as follows. Taking into account (\ref{s1s2}), we can write
\begin{eqnarray}
z^{r-p}&=&a^{[r-p]}(i+p)+\sum_{k\geq 1}q_k^{(n,r-p)}(i+p)z^{k(n-1)}a^{[r-p-kn]}(i+p)
\nonumber \\
&=&a^{[-p]}(i+p)\left(a^{[r]}(i)+\sum_{k\geq 1}q_k^{(n,r-p)}(i+p)z^{k(n-1)}a^{[r-kn]}(i)\right)
\nonumber
\end{eqnarray}
and
\[
z^{r-p}a^{[p]}(i)=a^{[r]}(i)+\sum_{k\geq 1}q_k^{(n,r-p)}(i+p)z^{k(n-1)}a^{[r-kn]}(i).
\]
Then writting explicitly the latter relation we get (\ref{integers}). Solving (\ref{integers}) in favor of $q_k^{(n,r-p)}(i+p)$ yields
\begin{equation}
a_k^{[p]}(i)+\sum_{j=1}^{k-1}q_j^{(n,r-(k-j)n)}(i)a_{k-j}^{[p]}(i)+q_k^{(n,r)}(i)=q_k^{(n,r-p)}(i+p).
\label{onegets}
\end{equation}

Observe that relations (\ref{Hps}) are coded in
\[
H^{(s)}=z^{s(1-n)}a^{[sn]}+\sum_{j=1}^sz^{(s-j)(1-n)}q_j^{(n,sn)}a^{[(s-j)n]}.
\]
This means that when restricting on ${\cal S}_0^n$, one has 
\begin{equation}
{\cal H}_{+}=<1, z^{1-n}a^{[n]}, z^{2(1-n)}a^{[2n]},...>.
\label{h+}
\end{equation}
We see that on ${\cal S}_0^n$ DKP chain equations can be written in the form of differential-difference conservation laws
\begin{equation}
\partial_s\xi_k(i)=F_k^{(n,s)}(i+1)-F_k^{(n,s)}(i).
\label{dKP}
\end{equation}
More exactly the equations (\ref{dKP}) appear as a result of projection of restricted DKP chain flows on the space ${\cal M}$ whose points are defined by infinite number of functions $\{a_k=a_k(i)\}$. As was shown in \cite{mpz}, with $n=1$, evolution equations (\ref{dKP}) are equivalent to the discrete KP hierarchy (dKP) \cite{ueno}. We can show that given any solution of DKP chain hierarchy restricted to ${\cal S}_0^n$, noninvertible map $g_n : a(i)\rightarrow z^{1-n}a^{[n]}(ni)$ gives solution of the dKP one. More generally, it is known that $g_k({\cal S}_{l-1}^{kn})\subset{\cal S}_{l-1}^n$ \cite{svinin2}. 

We refer, for simplicity, to equations (\ref{dKP}), with some fixed  as $n$th dKP hierarchy. Since all $n$th dKP hierarchy flows `live' on the same phase-space ${\cal M}$ it is natural to call (\ref{dKP}) (with any $n$) as extended dKP hierarchy \cite{svinin}. 

It is useful for our aims to write equations
\begin{eqnarray}
D_{t_s}q_k^{(n, r)}(i)&=&q_{k+s}^{(n, r)}(i+sn)+\sum_{j=1}^sq_j^{(n,sn)}(i)q_{k+s-j}^{(n, r)}(i+(s-j)n) \nonumber \\
& &-q_{k+s}^{(n, r)}(i)-\sum_{j=1}^sq_j^{(n,sn)}(i+r-(k+s-j)n)q_{k+s-j}^{(n, r)}(i) \label{Dts}
\end{eqnarray}
which are fulfilled in virtue of (\ref{dKP}). These equations can be found by making of use suitable differential-difference Lax equation (see, for example \cite{svinin2}).

\subsection{Restriction of DKP chain on ${\cal S}_0^n\cap{\cal S}_{l-1}^p$}

Let us now consider nontrivial intersections ${\cal S}_0^n\cap{\cal S}_{l-1}^p$. The word `nontrivial' means that $ln-p\neq 0$ to be supposed. According to the Theorem 1, the restriction of DKP chain on ${\cal S}_0^n\cap{\cal S}_{l-1}^p$ is defined by the generating relation
\[
z^{l-p}a^{[p]}-F^{(n,l)}-\sum_{j=1}^la_j^{[p]}F^{(n,l-j)}=\sum_{k\geq 1}J_k^lz^{-k}=0
\]
with Laurent series
\[
F^{(n,s)}\equiv z^s+\sum_{k\geq 1}F_k^{(n,s)}z^{-k}.
\]
Thus, restriction of DKP chain flows on ${\cal S}_0^n\cap{\cal S}_{l-1}^p$ is given by the following polynomial conditions:
\begin{equation}
J_k^l\equiv a^{[p]}_{k+l}-F_k^{(n,l)}-\sum_{j=1}^{l-1}a_j^{[p]}F_k^{(n,l-j)}=0,
\;\; \forall k\geq 1
\label{cond}
\end{equation}
which, in fact, define some submanifold ${\cal M}_{n,p,l}\subset{\cal M}$ invariant with respect to flows of $n$th discrete KP hierarchy (\ref{dKP}). Making of use relation (\ref{onegets}) we get the following 
\[
F_k^{(n,l)}(i)+\sum_{j=1}^{l-1}a_j^{[p]}(i)F_k^{(n,l-j)}(i)=a_{k+l}^{[ln]}(i)+\sum_{j=1}^{l-1}q_j^{(n,ln-p)}(i+p)a_{k+l-j}^{[(l-j)n]}(i)
\]
and, hence,
\[
J_k^l=a^{[p]}_{k+l}(i)-a_{k+l}^{[ln]}(i)-\sum_{j=1}^{l-1}q_j^{(n,ln-p)}(i+p)a_{k+l-j}^{[(l-j)n]}(i).
\]
Replacing in (\ref{integers}) $k\rightarrow k+l$ and setting $r=ln$ we get
\begin{equation}
J_k^l=Q_k^l+\sum_{j=1}^{k-1}a_j^{[-(k-j)n]}Q_{k-j}^l,\;\;\;
k\geq 1,
\label{JQ}
\end{equation}
with $Q_k^l(i)\equiv q_{k+l}^{(n,ln-p)}(i+p)$. Solving these relations in favor of $Q_k^l$ yields
\begin{equation}
Q_k^l=J_k^l+\sum_{j=1}^{k-1}q_j^{(n,-(k-j)n)}J_{k-j}^l,\;\;\;
k\geq 1.
\label{QJk}
\end{equation}
From (\ref{Dts}) we have   
\begin{eqnarray}
D_{t_s}Q_k^l(i)&=&
Q_{s+k}^l(i+sn)+\sum_{j=1}^sq_j^{(n, sn)}(i+p)Q_{s+k-j}^l(i+(s-j)n)
\nonumber \\
&&-Q_{s+k}^l(i)-\sum_{j=1}^sq_j^{(n, sn)}(i-(s+k-j)n)Q_{s+k-j}^l(i).
\nonumber
\end{eqnarray}
It is worthwhile to notice that the coefficients of this equation do not depend on $l$. The same is true for coefficients of transformation (\ref{QJk}). 

\subsection{Examples of integrable lattices}

Let us give below some examples of differential-difference equations which appear as a result of restricting of $n$th dKP hierarchy on ${\cal M}_{n,p,l}$. 

\begin{enumerate}
\item {\it One-field lattices}. 
The submanifold ${\cal M}_{n,p,1}$ is defined by infinite set of conditions $J_k^1=a_k^{[p]}-a_k^{[n]}=0,\;\;\; \forall k\geq 2$. Without loss of generality one can suppose that $p\geq 1$ and $n<p$ and $n\neq 0$. From (\ref{s1s21}) one has
\begin{eqnarray}
a_2^{[p]}(i)&=&a_2^{[n]}(i)+a_1^{[n]}(i)a_1^{[p-n]}(i+n)+a_2^{[p-n]}(i+n) \nonumber \\
&=&a_2^{[n]}(i+p-n)+a_1^{[n]}(i+p-n)a_1^{[p-n]}(i)+a_2^{[p-n]}(i) \nonumber
\end{eqnarray}
and
\[
a_2^{[p-n]}(i)=-a_1^{[n]}(i-n)a_1^{[p-n]}(i)
\]
and
\[
a_2^{[n]}(i+p-n)-a_2^{[n]}(i)=a_1^{[p-n]}(i)\left(a_1^{[n]}(i-n)-a_1^{[n]}(i+p-n)\right).
\]
From the latter one has following differential-difference equation\footnote{Here and in what follows $a_i\equiv a_1(i)$}
\begin{eqnarray}
\partial a_1^{[p-n]}(i)&=&a_i^{\prime}+\cdot\cdot\cdot +a_{i+p-n-1}^{\prime}=a_2^{[n]}(i+p-n)-a_2^{[n]}(i) \nonumber \\
&=&a_1^{[p-n]}(i)\left(a_1^{[n]}(i-n)-a_1^{[n]}(i+p-n)\right). \label{gen}
\end{eqnarray}
One needs to consider two cases.
Let $n\geq 1$ and $p\geq n+1$. Then (\ref{gen}) is specified as 
\begin{equation}
\sum_{s=1}^{p-n}a_{i+s-1}^{\prime}=\sum_{s=1}^{p-n}a_{i+s-1}\left(\sum_{s=1}^na_{i-s}-\sum_{s=p-n}^{p-1}a_{i+s}\right).
\label{Bll}
\end{equation}
Important case to consider is  $n\geq 1$ and $p= n+1$ which corresponds to Bogoyavlenskii lattice (\ref{Bl}). Let $n\leq -1$ and $p\geq 1$. In this case (\ref{gen}) becomes
\begin{equation}
\sum_{s=1}^{p+|n|}a_{i+s-1}^{\prime}=\sum_{s=1}^{p+|n|}a_{i+s-1}\left(\sum_{s=p+1}^{p+|n|}a_{i+s-1}-\sum_{s=1}^{|n|}a_{i+s-1}\right).
\label{np}
\end{equation}
It should be noted that two pairs of integers $(n, p)$ and $(-p, |n|)$ corresponds to the same equation of the form (\ref{np}).

\item {\it Toda lattice}. When restricting dKP hierarchy on ${\cal M}_{1,1,2}$, one requires  
\[
J_k^2=a_{k+2}(i)-a_{k+2}^{[2]}(i)-q_1^{(1,1)}(i+1)a_{k+1}(i)=0.
\]
The latter is solved by 
\[
a_{k+2}(i)=-\sum_{j=1}^ka_j(i-1)a_{k-j+2}(i).
\] 
Equations of the first flow of dKP hierarchy in this case are reduced to a pair of evolution equations 
\[
a_1^{\prime}(i)=a_2(i+1)-a_2(i),\;\;\;
a_2^{\prime}(i)=a_2(i)(a_1(i-1)-a_2(i))
\] 
which are equivalent to Toda lattice in its exponential form \cite{toda}

\item {\it Belov-Chaltikian lattice}. Restricting first flow of dKP hierarchy on ${\cal M}_{1,3,2}$ yields two-field system
\begin{eqnarray}
\displaystyle
\fl
a_1^{\prime}(i)&=&a_2(i+1)-a_2(i),
\nonumber \\
\fl
a_2^{\prime}(i)&=&a_1(i)a_2(i-1)-a_1(i-1)a_2(i+1)+a_2(i)\left\{a_1(i-2)+a_1(i-1)\right.
\nonumber \\
\fl
&&\left.-a_1(i)-a_1(i+1)\right\}+a_1(i-1)a_1(i)(a_1(i-2)-a_1(i+1))
\nonumber
\end{eqnarray}
wcich, in turn, via invertible ansatz
\[
L_i=-a_1(i),\;\;\;
W_i=a_2(i+1)+a_1(i)a_1(i+1)
\]
can be transformed into Belov-Chaltikian lattice \cite{belov}
\[
L^{\prime}_i=W_{i-1}-W_i+L_i(L_{i+1}-L_{i-1}),
\] 
\[
W^{\prime}_i=W_i(L_{i+2}-L_{i-1}).
\]

\item {\it Shabat dressing lattice}. Consider ${\cal M}_{1,0,2}$ which is defined by conditions 
\begin{equation}
J_k^2=-a_{k+2}^{[2]}-q_1^{(1,2)}a_{k+1}=0.
\label{conditi}
\end{equation}
One can check that in virtue of (\ref{conditi}) with $k=1$, 
\[
D_{x}q_2^{(1, 2)}(i)=D_x(-a_2(i)-a_2(i+1)+a_i^2)=0.
\]
So, we can write $a_2(i)+a_2(i+1)=a_i^2-\mu_i$, where $\mu_i$ do not depend on $x$ and
\[
a_i^{\prime}+a_{i+1}^{\prime}=a_2(i+2)-a_2(i)=a_{i+1}^2-a_i^2-\mu_{i+1}+\mu_i.
\]
The latter is nothing but Shabat dressing lattice \cite{shabat}.

\end{enumerate}

\section{Reductions of $n$th dKP hierarchy}
\label{sec:main}

The main goal of this section is to show some class of restrictions compatible with $n$th dKP hierarchy (\ref{dKP}). Corresponding constraints are supposed to be written in the form of periodicity conditions
\begin{equation}
I_k^l(i+n)=I_k^l(i),\;\;\;
\forall k\geq 1
\label{period}
\end{equation}
with suitable infinite collection of polynomial discrete functions $\{I_k^l=I_k^l : k\geq1\}$.
We are looking for these functions through invertible relations\footnote{In what follows we use simplified notations $I_k\equiv I_k^l$, $Q_k\equiv Q_k^l$ etc}
\begin{equation}
Q_k=I_k+\sum_{j=1}^{k-1}\zeta_{k-1,j}I_{k-j},\;\;\;
k\geq 1
\label{Q1}
\end{equation}
with some unknown coefficients $\zeta_{k,j}$. We have
\begin{eqnarray}
\displaystyle
\fl
D_{t_s}I_1(i)=D_{t_s}Q_1(i)&=&
Q_{s+1}(i+sn)+\sum_{j=1}^sq_j^{(n, sn)}(i+p)Q_{s-j+1}(i+(s-j)n)
\nonumber \\
\fl
&&-Q_{s+1}(i)-\sum_{k=1}^sq_j^{(n, sn)}(i-(s-j+1)n)Q_{s-j+1}(i).
\label{sI1}
\end{eqnarray}
Substituting (\ref{Q1}) into  right-hand side of (\ref{sI1}) we require that it is identically zero provided that the conditions (\ref{period}) are imposed. This leads to relations
\[
\fl
\zeta_{s,m}(i+sn)+\sum_{j=1}^{m-1}q_j^{(n,sn)}(i+p)\zeta_{s-j,m-j}(i+(s-j)n)+q_m^{(n,sn)}(i+p)
\]
\[
\fl
=\zeta_{s,m}(i)+\sum_{j=1}^{m-1}q_j^{(n,sn)}(i-(s-j+1)n)\zeta_{s-j,m-j}(i)+q_m^{(n,sn)}(i-(s-m+1)n)
\]
with $s\geq 1$ and $m=1,..., s$. Taking into account (\ref{r1r2}), one can easily check that the solution of these equations is given by
\begin{equation}
\zeta_{s,m}(i)=q_m^{(n,-p-(s-m+1)n)}(i+p).
\label{zeta}
\end{equation}
Thus, we have the following: if conditions (\ref{period}) for $I_k$ defined through the relations
\begin{equation}
Q_k(i)=I_k(i)+\sum_{j=1}^{k-1}q_j^{(n,-p-(k-j)n)}(i+p)I_{k-j}(i),\;\;\;
k\geq 1
\label{Q2}
\end{equation}
are valid then $D_{t_s}I_1=0,\;\; \forall s\geq 1$. Solving (\ref{Q2}) in favor of $I_k$ yields
\begin{equation}
I_k(i)=Q_k(i)+\sum_{j=1}^{k-1}a_j^{[-p-(k-j)n)]}(i+p)Q_{k-j}(i),\;\;\;
k\geq 1.
\label{Q3}
\end{equation}

We can prove, by induction, that provided (\ref{period}) the quantities $I_k$ do not depend on evolution parameters $t_s$ for all $k\geq 1$. We have
\[
D_{t_s}I_{k+1}(i)+\sum_{j=1}^kD_{t_s}\zeta_{k,j}(i)I_{k-j+1}(i)+\sum_{j=1}^k\zeta_{k,j}(i)D_{t_s}I_{k-j+1}(i)=D_{t_s}Q_{k+1}(i)
\]
\[
=Q_{s+k+1}(i+sn)+\sum_{j=1}^sq_j^{(n, sn)}(i+p)Q_{s-j+k+1}(i+(s-j)n)
\]
\[
-Q_{s+k+1}(i)-\sum_{j=1}^sq_j^{(n, sn)}(i-(s-j+k+1)n)Q_{s-j+k+1}(i).
\]
Let us suppose that we already proved that $D_{t_s}I_j$ for $j=1,..., k$ in virtue of (\ref{period}) with $I_k$ given by (\ref{Q2}). Then $D_{t_s}I_{k+1}=0$ under (\ref{period}) if the relations
\[
\fl
\zeta_{s+k,m}(i+sn)+\sum_{j=1}^{m-1}q_j^{(n,sn)}(i+p)\zeta_{s+k-j,m-j}(i+(s-j)n)+q_m^{(n,sn)}(i+p)
\]
\[
\fl
=\zeta_{s+k,m}(i)+\sum_{j=1}^{m-1}q_j^{(n,sn)}(i-(s+k-j+1)n)\zeta_{s+k-j,m-j}(i)+q_m^{(n,sn)}(i-(s+k-m+1)n)
\]
with $m=1,..., s$ and
\[
\fl
\zeta_{s+k,s+m}(i+sn)+\sum_{j=1}^sq_j^{(n,sn)}(i+p)\zeta_{s+k-j,s+m-j}(i+(s-j)n)
\]
\[
\fl
=\zeta_{s+k,s+m}(i)+\sum_{j=1}^sq_j^{(n,sn)}(i+(s+k-j+1)n))\zeta_{s+k-j,s+m-j}(i)+D_{t_s}\zeta_{k,m}(i)
\]
with $m=1,..., k$ are valid. Again we can check that these relations are solved by (\ref{zeta}). As an obvious consequence of the above calculations we obtain following theorem:

{\bf Theorem 3.} Periodicity conditions (\ref{period}) with $I_k^l$ given by (\ref{Q3}) are compatible with $n$th discrete KP hierarchy.

Let us denote the submanifold of ${\cal M}$ defined by conditions of periodicity (\ref{period}) as ${\cal N}_{n,p,l}$. Infinite set of constraints $I_k^l=0$ defining ${\cal M}_{n,p,l}$ give particular solution of (\ref{period}) and, hence, we have ${\cal M}_{n,p,l}\subset{\cal N}_{n,p,l}$. 

It is useful to establish relationship between $J_k^l$ and $I_k^l$. Making of use (\ref{integers}), (\ref{JQ}) and (\ref{Q2}), we get the relation
\[
J_k=I_k+\sum_{j=1}^{k-1}a_j^{[p]}I_{k-j}
\] 
and its inverse
\[
I_k(i)=J_k(i)+\sum_{j=1}^{k-1}a_j^{[-p]}(i+p)J_{k-j}(i).
\]

\section{Reductions of Bogoyavlenskii lattice}
\label{sec:Bl}
 
The goal of this Section is to show how Theorem 3 can be applied for constructing of some class of constraints compatible with Bogoyavlenskii lattice (\ref{Bl}).  More exactly, we would like to show a class of restrictions which correspond to intersection ${\cal M}_{n,n+1,1}\cap{\cal N}_{n,p,1}$ which is evidently equivalent to ${\cal M}_{n+1,n,1}\cap{\cal N}_{n+1,p,1}$. The submanifold ${\cal M}_{n,n+1,1}$ is defined by equations $a^{[n]}_k=a^{[n+1]}_k$ with $k\geq 2$ which uniquely solved as $a_k=P^{[n]}_k[a]$ with some discrete polynomials $P^{[n]}_k$. Next we require that the periodicity conditions $I^1_k(i+n)=I^1_k(i)$ and $I^1_k(i+n+1)=I^1_k(i)$ must be valid simutaneously. This is equivalent to $I^1_k(i+1)=I^1_k(i)$. 
It is natural to consider three different cases:

\begin{enumerate}
\item 
Let $p\geq n+2$. Keeping in mind the condition $a_2^{[n]}=a_2^{[n+1]}$, one calculate to obtain
\begin{eqnarray}
I_1^1(i)&=&J_1^1(i)=a_2^{[p]}(i)-a_2^{[n]}(i)=a_i\left(a_{i+n+1}+\cdot\cdot\cdot +a_{i+p-1}\right)
\nonumber \\
&&+a_{i+1}\left(a_{i+n+2}+\cdot\cdot\cdot+a_{i+p-1}\right)+\cdot\cdot\cdot +a_{i+p-n-2}a_{i+p-1}.
\nonumber
\end{eqnarray}
The condition $I_1^1(i+1)=I_1^1(i)$ can be written as
\begin{equation}
a_{i+p}=a_i\frac{Q_{i+n}}{Q_i},\;\;\mbox{with}\;\; 
Q_i\equiv\sum_{j=1}^{p-n-1}a_{i+j}. 
\label{difference}
\end{equation}
It is natural to suppose that this difference equation has a number of integrals enough for its integrability \cite{veselov}. If so, then this collection of integrals is provided by $\{I_k^1\}$ but one must to calculate its explicit form $I_k^1=I_k^1[a]$ in each case. Neverthless, we can write explicit form of two integrals for (\ref{difference}):
\[
K_i=\frac{\prod_{s=1}^pa_{i+s-1}} {\prod_{j=1}^nQ_{i+j-1}}\;\;\mbox{and}\;\; P_i=\frac{\prod_{j=1}^{n+1}\left(a_{i+j-1}+Q_{i+j-1}\right)} {\prod_{j=1}^nQ_{i+j-1}}.
\]
which directly follows from the form of this equation. Second integral here is derived as follows. As a consequence of (\ref{difference}) we can write 
\[
a_{i+p}\left(a_i+Q_i\right)=a_i\left(a_{i+n+1}+Q_{i+n+1}\right)
\] 
and 
\[
Q_{i+n}\left(a_i+Q_i\right)=Q_i\left(a_{i+n+1}+Q_{i+n+1}\right).
\]
It is obvious that the latter relation can be written as $P_i=P_{i+1}$.

\item
Let $p\leq -1$. We calculate to write down
\[
\fl
I_1^1(i)=a_{i-1}\left(a_{i-|p|}+\cdot\cdot\cdot +a_{i+n-1}\right)
+a_{i-2}\left(a_{i-|p|}+\cdot\cdot\cdot +a_{i+n-2}\right)+\cdot\cdot\cdot +a_{i-|p|-n}a_{i-|p|}.
\]
The corresponding constraint can be written as
\[
a_{i+|p|+n}Q_{i+n}=a_{i+n}Q_{i-1},\;\;\mbox{with}\;\; Q_i\equiv\sum_{j=1}^{|p|+n}a_{i+j}.
\]
Solving this relation in favor of $a_{i+|p|+2n}$ yields
\begin{equation}
a_{i+2n+|p|}=\frac{a_{i+n}}{a_{i+|p|+n}}Q_{i-1}-\sum_{j=n+1}^{2n+|p|-1}a_{i+j}.
\label{yield}
\end{equation}
In this case we also able to write two integrals for (\ref{yield}) in its explicit form
\[
K_i=\prod_{j=1}^{|p|}a_{i+j-1}\prod_{j=1}^{n+1}Q_{i-j}\;\;\mbox{and}\;\; 
P_i=\frac{\prod_{j=1}^n\left(a_{i-j}+Q_{i-j}\right)}{\prod_{j=1}^{n+1}Q_{i-j}}.
\]

\item
Let $p=1,..., n-1$. Then
\[
\fl
I_1^1(i)=a_{i-1}\left(a_{i+p}+\cdot\cdot\cdot +a_{i+n-1}\right)+a_{i-2}\left(a_{i+p}+\cdot\cdot\cdot +a_{i+n-2}\right)+\cdot\cdot\cdot
+a_{i+p-n}a_{i+p}.
\]
We can write down corresponding constraint in the form
\[
a_{i+n-p}Q_{i+n}=a_{i+n}Q_{i-1},\;\;\mbox{with}\;\; Q_i\equiv\sum_{j=1}^{n-p}a_{i+j}
\]
and 
\[
a_{i+2n-p}=\frac{a_{i+n}}{a_{i+n-p}}Q_{i-1}-\sum_{j=n+1}^{2n-p-1}a_{i+j}.
\]
Two integrals $K$ and $P$ are:
\[
K_i=\frac{\prod_{j=1}^pa_{i-j}} {\prod_{j=1}^{n+1}Q_{i-j}}\;\;\mbox{and}\;\;
P_i=\frac{\prod_{j=1}^n\left(a_{i-j}+Q_{i-j}\right)} {\prod_{j=1}^{n+1}
Q_{i-s}}.
\]
\end{enumerate}

One sees that in each of three cases corresponding constraints are written in the form of ordinary difference equation $a_{i+N}=R(a_i,..., a_{i+N-1})$ with $N=p$, $N=2n+|p|$ and $N=2n-p$, respectively. 

Finally, let us show simple example corresponding to $p=4$ and $n=1$, that is, when Volterra lattice (\ref{Vl}) is constrained by condition given by fourth-order difference equation  
\[
a_{i+4}=a_i\frac{a_{i+2}+a_{i+3}}{a_{i+1}+a_{i+2}}.
\]
with three integrals
\[
K_i=\frac{a_ia_{i+1}a_{i+2}a_{i+3}}{a_{i+1}+a_{i+2}},\;\;\;
P_i=\frac{(a_i+a_{i+1}+a_{i+2})(a_{i+1}+a_{i+2}+a_{i+3})}{a_{i+1}+a_{i+2}}.
\]
\[
J_i=a_i\left(a_{i+2}+a_{i+3}\right)+a_{i+1}a_{i+3},
\]
Observe that $I_1^1=J$ and $I_2^1=-K-JP$. Identify now $a_i=u_1, a_{i+1}=u_2, a_{i+2}=u_3, a_{i+3}=u_4$. Attached system of ordinary differential equations looks as follows:
\[
u_1^{\prime}=u_1\left(u_4\frac{u_1+u_2}{u_2+u_3}-u_2\right),\;\;\;
u_2^{\prime}=u_2\left(u_1-u_3\right),\;\;\;
u_3^{\prime}=u_3\left(u_2-u_4\right),\;\;\;
\]
\[
u_4^{\prime}=u_4\left(u_3-u_1\frac{u_3+u_4}{u_2+u_3}\right),
\]
This system has three first integrals 
\[
K=\frac{u_1u_2u_3u_4}{u_2+u_3},\;\;\;
P=\frac{(u_1+u_2+u_3)(u_2+u_3+u_4)}{u_2+u_3}.
\]
\[
J=u_1\left(u_3+u_4\right)+u_2u_4.
\]
Shifting $i\rightarrow i+1$ gives discrete symmetry transformation
\begin{equation}
\bar{u}_k=u_{k+1},\;\;\;
k=1, 2, 3,\;\;\;
\bar{u}_4=u_1\frac{u_3+u_4}{u_2+u_3}
\label{symm}
\end{equation}
By direct calculations one can check that integrals $K, P$ and $J$ are invariant under (\ref{symm}).

\section{Conclusion}

We have presented a scheme for constructing a broad class of constraints compatible with integrable lattices which can be derived as reductions of DKP chain. For particular case of the Bogoyavlenskii lattice we showed some simple examples in its explicit form. In principle, the solution of the lattice under consideration constrained by some condition is constructed as follows. One solves attached system of ordinary differential equations with some initial conditions $(u_1^0,..., u_N^0)\in{\bf R}^N$ to obtain 
\[
a(i, x)=u_1(x),..., a(i+N-1, x)=u_N(x)
\] 
for some initial value, $i=i_0$. To find $a(i, x)$ for remaining values of discrete variable $i$ one needs to use discrete symmetry transformation. 

In this connection, it is important to determine the structure of solutions for attached systems of ordinary differential equations. We suppose that all these systems are integrable in  Liouville sense with first integrals invariant with respect to discrete symmetry transformation generated by shifting $i\rightarrow i+1$.

\end{document}